\input{aipcheck}

\documentclass[
    ,final            
  ]
  {aipproc}

\layoutstyle{6x9}

\def\Msun {M$_{\scriptscriptstyle \odot}$}
\def \ltaprx {\lower .1ex\hbox{\rlap{\raise .6ex\hbox{\hskip .3ex
	{\ifmmode{\scriptscriptstyle <}\else 
		{$\scriptscriptstyle <$}\fi}}}
	\kern -.4ex{\ifmmode{\scriptscriptstyle \sim}\else 
		{$\scriptscriptstyle\sim$}\fi}}}
\def\gtaprx {\lower .1ex\hbox{\rlap{\raise .6ex\hbox{\hskip .3ex
	{\ifmmode{\scriptscriptstyle >}\else 
		{$\scriptscriptstyle >$}\fi}}}
	\kern -.4ex{\ifmmode{\scriptscriptstyle \sim}\else 
		{$\scriptscriptstyle\sim$}\fi}}}

\begin{document}

\title{The Supernova Gamma-Ray Burst Connection}

\classification{98.70.Rz,97.60.Bw,26.50+x}
\keywords      {supernovae,gamma-ray bursts,stellar evolution}

\author{S. E. Woosley}{
  address={Department of Astronomy and Astrophysics, UCSC, Santa Cruz CA 95064}
}

\author{A. Heger$^*$}{
  address={Theoretical Astrophysics Group, T-6, MS B227, Los Alamos National
Laboratory, Los Alamos, NM 87545}
}

\begin{abstract}
The chief distinction between ordinary supernovae and long-soft
gamma-ray bursts (GRBs) is the degree of differential rotation in the
inner several solar masses when a massive star dies, and GRBs are rare
mainly because of the difficulty achieving the necessary high rotation
rate. Models that do provide the necessary angular momentum are
discussed, with emphasis on a new single star model whose rapid
rotation leads to complete mixing on the main sequence and avoids red
giant formation. This channel of progenitor evolution also gives a
broader range of masses than previous models, and allows the copious
production of bursts outside of binaries and at high redshifts.
However, even the production of a bare helium core rotating nearly at
break up is not, by itself, a sufficient condition to make a gamma-ray
burst. Wolf-Rayet mass loss must be low, and will be low in regions of
low metallicity. This suggests that bursts at high redshift (low
metallicity) will, on the average, be more energetic, have more time
structure, and last longer than bursts nearby. Every burst consists of
three components: a polar jet ($\sim$0.1 radian), high energy,
subrelativistic mass ejection ($\sim$1 radian), and low velocity
equatorial mass that can fall back after the initial explosion. The
relative proportions of these three components can give a diverse
assortment of supernovae and high energy transients whose properties
may vary with redshift.
\end{abstract}

\maketitle

\section{Introduction}

As talks summarized elsewhere in this proceedings and papers in the
literature have made clear \citep{Woo06a}, most GRBs of the long-soft
variety (henceforth just GRBs) are a consequence of the deaths of
massive stars. Evidence supporting this comes from: 1) the location of
GRBs in regions of active star formation \citep{Fru06}; 2) the clear
presence of supernovae of Type Ic-BL (``broad-lined Ic'') in
conjunction with three GRBs: 980425, 030329, and 031203; 3) the
presence of supernova-like bumps in most other GRBs where they might
be observed \citep{Zeh04}; and 4) the similarity in energy between the
beaming-corrected GRB energy and that of a supernova.

Accepting this as a starting point, an important question must be why
some massive stars die as supernovae of the ordinary variety, while
others die as GRBs. The fraction that do die as GRBs is apparently very
small. Taking an event rate of core-collapse supernovae visible from
the earth today as $\sim$20 per 16 arc min squared per year
\citep{Mad98}, the integrated rate of supernovae on the sky is about 6
per second. BATSE saw about a burst a day. Correcting by a factor of
300 for beaming and another factor of three for Earth occultation and
bursts that were missed for reasons other than beaming, the GRB sky
rate is about 0.02 per second.  That is, GRBs are a fraction of order
0.3\% of all massive star deaths. This fraction could be larger if
there are numerous sub-luminous events like GRB 980425 or cosmic X-ray
flashes (XRFs), and it might increase with redshift, but apparently
GRBs are a rare channel of massive star death.

Another interesting question is whether GRBs and ordinary supernovae
are the extrema of a continuum of events or separate classes of
explosions.  The existence of such diverse phenomena as XRFs, 980425,
and GRBs with varying energy and supernova properties suggests a
continuum. So, too, does the growing class of ``hyper-energetic'' or
grossly asymmetric supernovae like SN 2005bf
\citep{Fol06,Tom05,Anu05}. Add to this the fact that most broad-lined
Type Ic supernovae do \emph{not} harbor GRBs \citep{Sod05}, and a
compelling case might be made for some continuously variable parameter
that dials between ordinary supernovae and energetic GRBs like 990123.
Clearly, the difference is that GRBs concentrate a significant
fraction of their 10$^{51}$ - 10$^{52}$ ergs into highly relativistic
($\Gamma > 200$) beamed ejecta while ordinary supernovae do not, but
searches beyond that for a deeper physical cause take us into what is
suspected, and away from what is certain.

\section{GRB Progenitors}

The two key quantities that determine how a massive star dies are its
mass and the rotation rate of its inner few solar masses. Stars that
are more massive when they die are more likely to make black
holes. More massive stars also evolve more quickly and may more
effectively preserve the large angular momentum they have at birth in
their core \citep{Heg05}. Rotation is a key ingredient in any
successful GRB model. If the rotational energy of a neutron star is to
provide the $\sim$10$^{52}$ erg inferred for some of the supernovae
accompanying GRBs, it must have a rotational period $\sim$1 ms. This
implies a specific equatorial angular momentum, $j \approx 7 \times
10^{15}$ cm$^2$ s$^{-1} \ (P/1 \ {\rm ms})(R/10 \ {\rm km})^2$, about
20 or more times that of a typical pulsar. If a disk is to form around
a black hole one needs more.  At 3 \Msun, the specific angular
momentum of the last stable orbit is $j_{lso} = 2 \sqrt{3} GM/c = 4.6
\times 10^{16} $ cm$^2$
s$^{-1}\ (M_{\rm BH}/3 M_{\scriptscriptstyle \odot})$ for a
Schwarzschild black hole, and $j_{lso} = 2/\!\sqrt{3} GM/c = 1.5
\times 10^{16}$ cm$^2$
s$^{-1}\ (M_{\rm BH}/3 M_{\scriptscriptstyle \odot})$ for an extreme
Kerr hole. Given that angular momentum increases monotonically with
interior mass from 1.5 to 3 \Msun, the necessary values for neutron
star models and black hole models are qualitatively similar.

Studies of massive star evolution that include estimated magnetic
torques \citep{Heg05} have shown that such large values of angular
momentum are not easily achieved if the star spends much time as a red
giant, or even if it loses its envelope early on, but has a high mass
loss rate afterwords as a Wolf-Rayet (WR) star \citep{HW03}.  A
frequently discussed scenario is a stellar merger, but the central
region of a differentially rotating star that has lost its angular
momentum is virtually impossible to spin up again. This limits
viable models to those where the merger removes the envelope (with
enough time still left for for its dispersal), produces a compact
helium star that never becomes a blue or red giant itself, and still
does not slow down the denser, inner regions of the star.

Several possibilities remain, however. One is that the magnetic
torques used in the above study were simply too large, but then one
must take care both to produce the large number of slowly rotating
neutron stars that are seen, and not overproduce GRBs. Another
possibility is a binary merger between two massive stars, both of
which are burning helium in their centers \citep{Fry05}. The merger
ejects both envelopes and the (now rapidly rotating) residual helium
star evolves to produce the burst a few hundred thousand years later,
after the envelope has left the vicinity. Even then, though, the new
helium star must lose a very limited amount of mass before dying, so
it helps if the merger occurs late during helium burning for one or
both stars.

A third possibility is that GRBs come from the merger of a black hole
or neutron star with either the helium core of a massive star
\citep{Fry98,Zha01} or a white dwarf \citep{Fry99}. In the former
case, the giant star's envelope must be completely dispersed before
the final merger occurs. In the white dwarf case, one would not expect
such a tight correlation with star forming regions, but a fraction of
GRBs could still be made this way. In both of these compact merger
models, however, there is so \emph{much} angular momentum that the
total duration of the event is quite long, longer than typical
GRBs. It could be that the observed burst only reflects the epoch of
maximum accretion in which case both enduring activity and long
precursors might be present. Or perhaps this model only makes very
long GRBs.

A final possibility, and one that has received a lot of attention
lately, is that GRBs result from a rare channel of single star
evolution in which red giant formation is avoided altogether. The
possibilities here are exciting and warrant a more lengthy discussion.

\subsection{A Single Star Model}

It has been recently realized \citep{Woo06b,Yoo06} that stars which
rotate \emph{very} rapidly on the main sequence, with equatorial
speeds near 400 km s$^{-1}$ instead of 200 - 300 km s$^{-1}$, may
experience nearly complete mixing on the main sequence
\citep{FL95,HL00}. In fact, rotationally-induced mixing in such stars,
chiefly by Eddington-Sweet circulation, keeps the star's composition
nearly homogeneous until the end of helium burning. A red giant is
never formed and the star goes straight from the main sequence to
being a Wolf-Rayet star. Such stars will have more rapidly rotating
iron cores when they die, though still not fast enough to be GRBs if
the helium star loses a lot of mass. Because the progenitors are on
the very high end of the observed distribution function for O- and
B-star rotation
\citep{Gie04}, GRBs will be a rare channel of star death.
 
\begin{figure}
\includegraphics[height=.4\textheight]{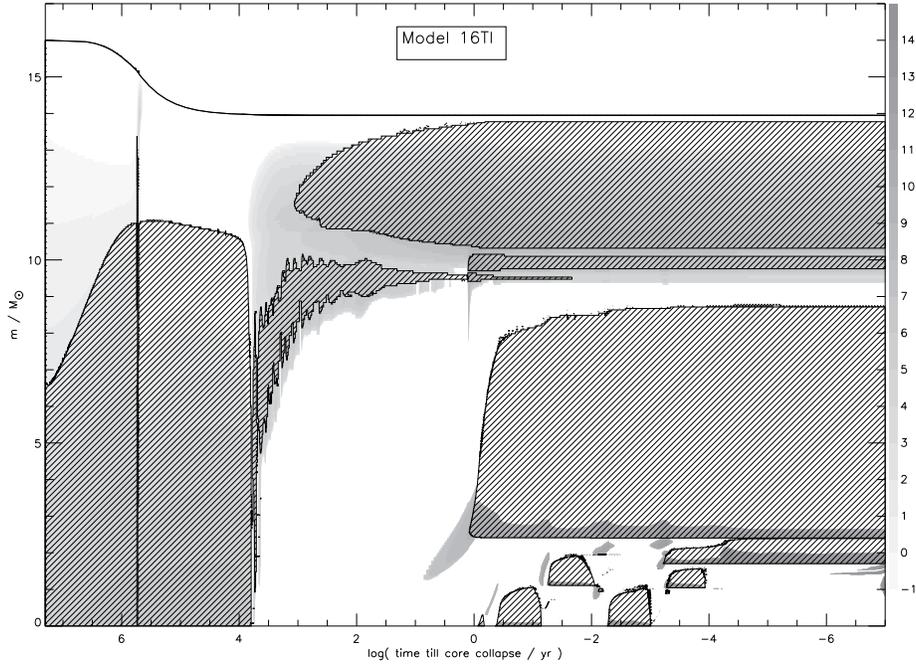}
\caption{Convective history of a rapidly rotating 16 \Msun \ star that
experiences nearly complete mixing on the main sequence and avoids red
giant formation. The left axis is the interior mass, cross hatching
shows regions that are convective, and gray shading indicates specific
nuclear energy generation on a logarithmic scale (right gray-scale bar, in
log ergs/g/s).  The bottom axis is the time until death (on a
logarithmic scale).  This model had an initial composition of one per
cent solar and an equatorial velocity half way through hydrogen
burning of 380 km s$^{-1}$. The final mass was 14.0 \Msun \ and the
iron core mass was 1.60 \Msun. The abundances in the surface
convection zone of the presupernova star were 9.5\% He, 30\% C, 57\%
O, and 2.7\% Ne. The radius was $4.1 \times 10^{10}$ cm. }
\end{figure}

This model has several beneficial features for GRBs. First, the
progenitor star is a compact WR star, mostly composed of oxygen with
little helium at its surface - a WO star. This agrees well with the
properties of those few GRB-supernovae that have been
well studied spectroscopically. Second, it is possible to produce a
wider range of GRB progenitor masses. Main sequence stars as light as
10 \Msun \ produce helium and heavy element cores that - \emph{modulo}
the mass loss - are about as large as that of a 25 \Msun \ star with
slow rotation (both about 9 \Msun). Such big cores evolve rapidly,
retain their angular momentum, and develop big iron cores when they
die. They are more likely to make black holes. Below about 10
\Msun, larger, possibly unphysical rotation speeds are
necessary on the main sequence to cause Eddington-Sweet circulation to
mix the star efficiently. On the upper end, very massive GRB
progenitors can be produced from main sequence stars of more moderate
mass than previously thought. For slowly rotating, solar-metallicity
stars, the largest helium core that can exist when the star dies is
around 15 \Msun, corresponding to a main sequence star of 35
\Msun. Heavier stars lose their envelopes to winds and the resulting
helium core shrinks by mass loss. But for these rapidly rotating,
well-mixed stars, the upper bound on the helium core is, in principle,
equal to the main sequence mass. Of course, mass loss, both on the main
sequence and especially as a WR star, will still shrink the mass. As
we shall see shortly however, turning the metallicity down can
alleviate the mass loss of a WR star, so that GRB progenitors in low
metallicity regions could have very big mass.

An upper limit to the helium core mass that comes from these rapidly
rotating models is the first mass to encounter the pulsational pair
instability \citep{Heg02}, about 40 \Msun \ of helium and heavy
elements. The evolved star experiences violent, repeated,
nuclear-powered explosions when the star ignites oxygen burning.  Each
outburst ejects solar masses of surface material. This material
surrounds the star when it finally dies (typically death happens
months to years later) and prevents a GRB from getting out (though
such explosions are interesting in their own right). For helium cores
above about 65 \Msun, the pair-instability becomes so violent that it
leads to the complete disruption of the star and no GRB will be made,
but above 140 \Msun \ a new regime is encountered where black holes
are formed and GRBs of a more energetic, longer-lasting variety become
possible \citep{Fry01}.  In slowly rotating stars, without mass loss,
the pulsational pair instability is first encountered for main
sequence stars of $\sim$100 \Msun \ and black holes are made starting
at about 260 \Msun, so these lower values for rapidly rotating stars
are very significant changes. The numbers are currently uncertain,
however, because few rapidly rotating models have been calculated
and the effect of the rotation on the pair instability has not been
included in a self-consistent way.

\begin{figure}
\begin{minipage}[b]{0.7\linewidth}
\centering
\includegraphics[width=\textwidth]{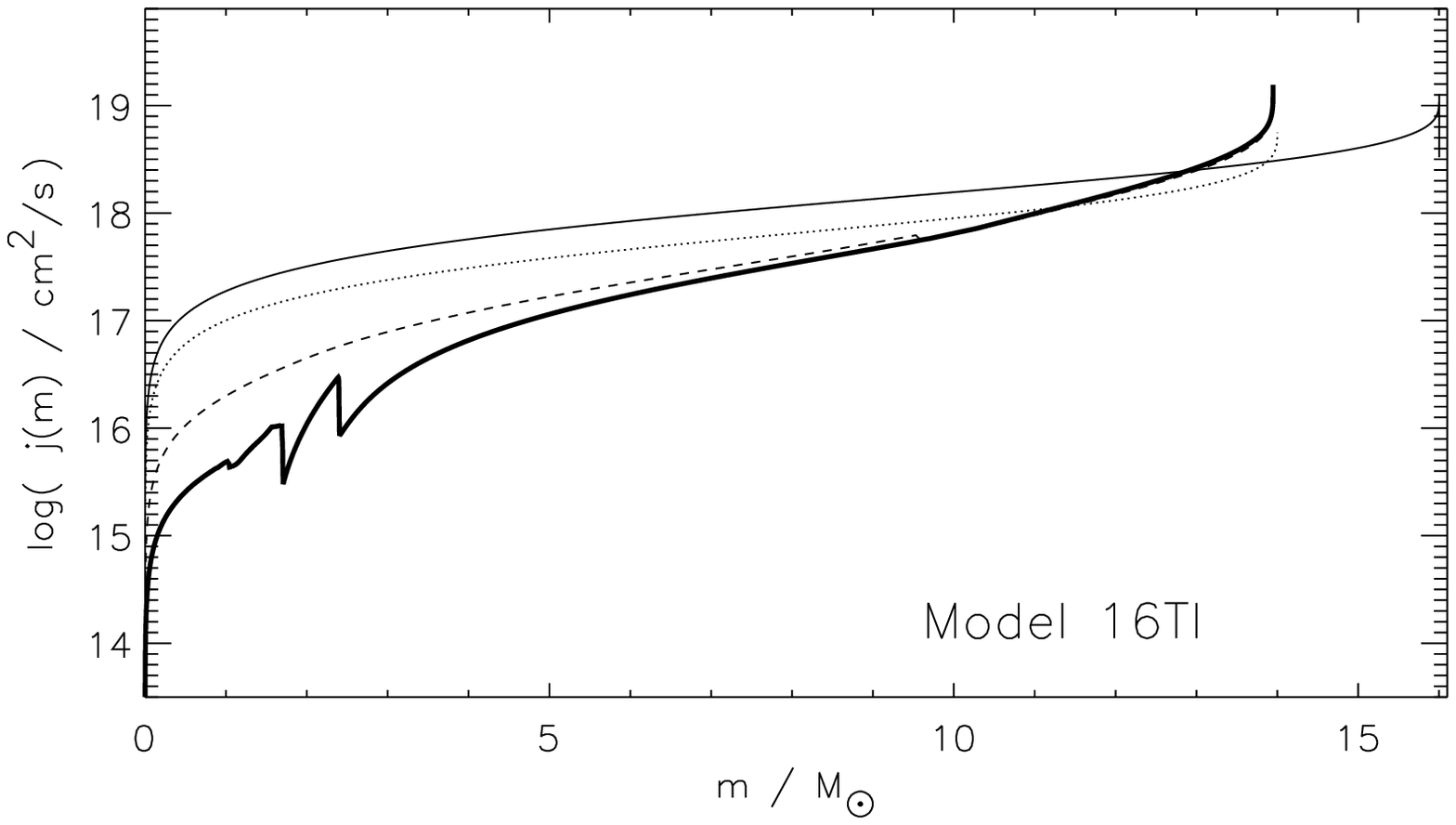}  
\includegraphics[width=\textwidth]{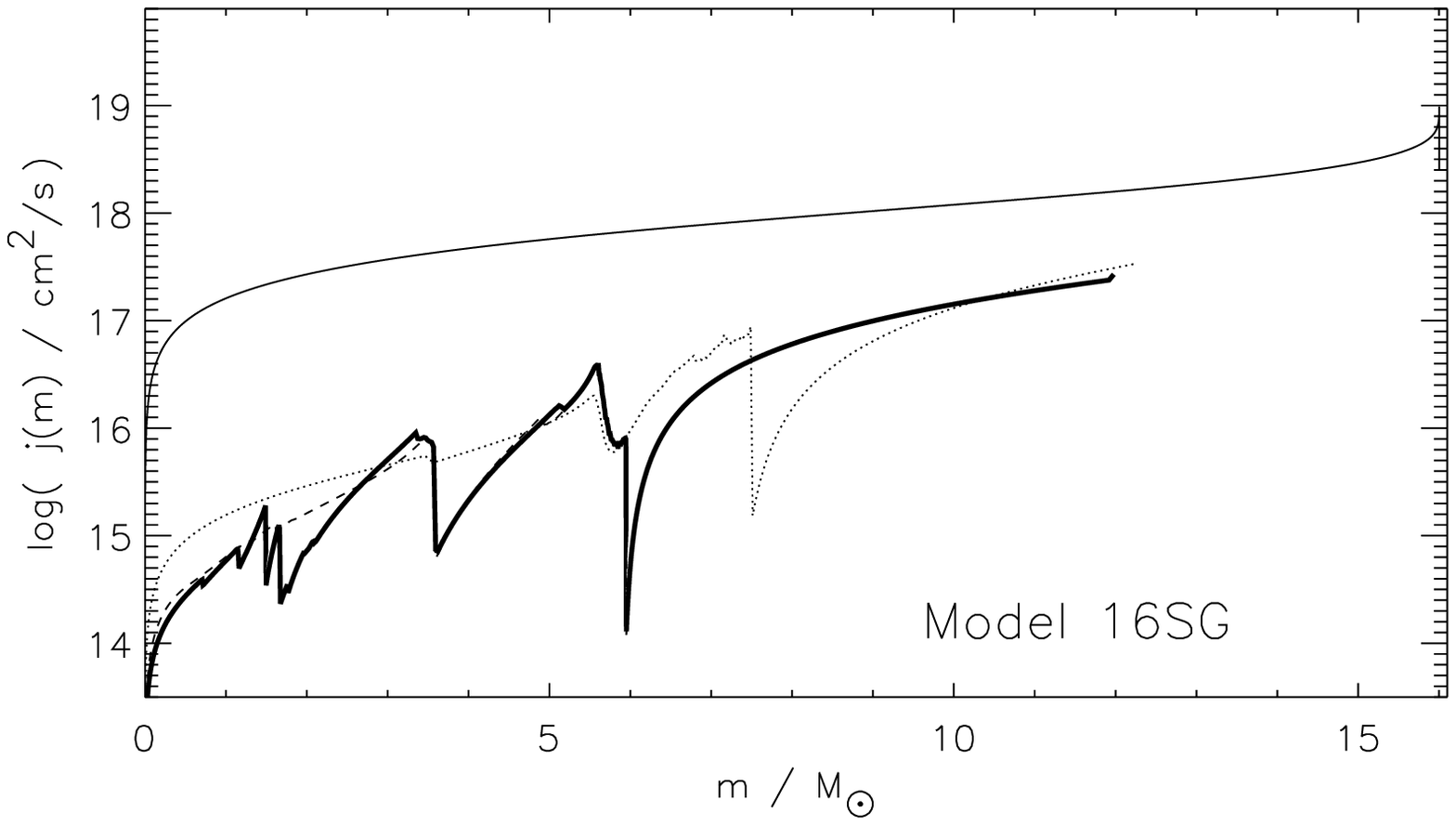}  
\end{minipage}
\caption{Specific angular momentum as a function of mass in two stars
with initial mass 16 \Msun \ \citep{Woo06b}. Both models included
angular momentum transport due to rotationally induced mixing
processes and magnetic torques. The angular momentum is evaluated at
the zero age main sequence (top line, solid), central helium depletion
(dotted line, next down), carbon depletion (dashed line), and
presupernova (dark solid line). The top star had an equatorial
rotation rate on the main sequence of 380 km s$^{-1}$, experienced
complete mixing, and avoided ever making a red giant.  The top star
also had a Wolf-Rayet stellar mass loss rate corresponding to 1\%
solar metallicity. The bottom star had solar metallicity and a slower
rotation rate on the main sequence, 215 km s$^{-1}$. The bottom star
became a red supergiant, lost more mass, and produced a slowly
rotating iron core that would give a 8 ms pulsar period after collapse
and neutrino loss. The top star ended up with a higher mass and much
greater rotation rate. This star would form a black hole accretion
disk starting at 3.5 \Msun. Avoiding red giant formation and reducing
the mass loss greatly increases the probability of making a GRB.}
\end{figure}

A final interesting consequence of rapid rotation and nearly
homogeneous mixing, is that single stars, even at low metallicity, can
end their lives as compact WR stars without the need of a binary
companion to absorb the envelope. This has interesting implications
for making GRBs at high redshift.

\subsection{The Metallicity Dependence of GRBs}

Whether the WR-star that serves as the progenitor of a GRB is made by
merger or rotationally-induced mixing, it is essential that its mass
loss, or more specifically, its angular momentum loss, be low. A
massive WR star typically spends 0.5 to 1 million years burning
helium.  If, during that time, its mass loss rate is over 10$^{-5}$
\Msun \ yr$^{-1}$, it will lose a significant fraction of its
mass. Magnetic torques maintain nearly rigid rotation during helium
burning, so the continued expansion of layers deep in the star to
larger radius brakes the rotation of the inner core beyond what is
needed later to make a GRB. Mass loss is the enemy of GRBs.

Fortunately, it has been recently determined \citep{Vin05} that
reducing the metal content of a WR star even by a factor of a few
lowers its mass loss very appreciably, approximately as
Z$^{0.86}$. Once the metallicity has been reduced by a factor of 10,
typical mass loss rates for 10 - 20 \Msun \ WR stars are
$\sim$10$^{-6}$ \Msun \ yr$^{-1}$ and less. Losing less than a solar
mass would not greatly alter the final angular momentum distribution
of a massive WR star, so at metallicities $\sim$10\% solar and less,
GRBs should be plentiful.  It is important to note that the
metallicity employed in the scaling here is the initial concentration
of iron in the star. It does not include the carbon and oxygen at the
surface of a WC or WO star that was made in the star itself during
helium burning, at least not until the iron abundance becomes so low
that the mass loss is negligible anyway, roughly 0.01 solar.

This does not mean that GRBs cannot happen in regions with solar
metallicity, only that it is harder. Measured WR mass loss rates show a
considerable spread and some merger models might also still work.  The
magnetic torques estimated in the models are uncertain and, for stars
that are highly deformed the angular dependence of the mass loss rate
is an issue. Mass loss preferentially from the poles would reduce the
amount of angular momentum carried away by each gram \citep{MM00}.

However, the key role of mass loss and its strong metallicity
dependence \emph{does} suggest that the fraction of massive stars dying
while making GRBs may be larger in regions where less nucleosynthesis
has occurred. As the metallicity declines it may also be possible to
make more energetic, longer lasting bursts.  Higher mass and angular
momentum at death increase the reservoir of material that can accrete
into a black hole. It may also lead to more rapidly rotating neutron
stars in the millisecond magnetar model.

The low mass loss rate also has implications for the afterglow
analysis, and a lower density may be more consistent with
observations \citep{Che04}. One must take care, however, since the wind
that is sampled in the afterglows was ejected during the post-helium
burning evolution of the star, during which the mass loss rate may have
varied from what is observed on the helium-burning main sequence.

\section{The Three Components of a GRB}

A GRB with an energetic supernova accompanying it will have three
components: 1) a highly relativistic, $\Gamma \gtaprx 200$, central
jet with an opening angle $\sim$0.1 radian measured from the
rotational axis; 2) a broader region of very energetic, but
subrelativistic ejecta extending out to angles $\sim$1 radian; and 3)
slower moving ejecta in the equator. In general, region 1 is
responsible for the GRB, region 2 is necessary for the supernova and
the $^{56}$Ni to make it bright, and region 3 is naturally present in
any model where outflow in the equator is blocked, e.g., by an
accretion disk in the collapsar model. Region 2 probably contains most
of the energy and is necessary because a narrow jet, by itself, is an
ineffective way of blowing up a star or producing explosive
nucleosynthesis. Region 3 may move out initially, but is largely
responsible for the ``fall-back'' that may keep an accretion-powered
source active long after the initial burst is over.  One might also
properly discuss a fourth region, the jet cocoon lying between
regions 1 and 2, but here we count that as part of the jet.

\begin{figure}
\centering
\includegraphics[height=.6\textheight]{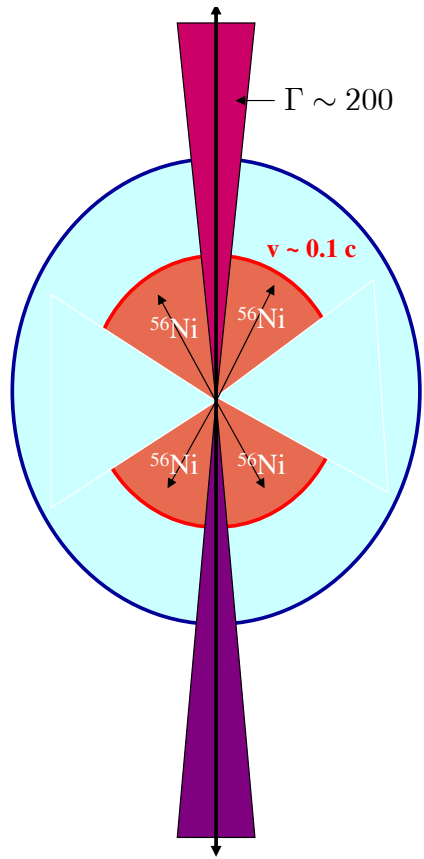}
\caption{Schematic illustrating the three components of a typical GRB
and its accompanying supernova. A core relativistic jet ($\sim$0.1
radian, $\Gamma \sim 200$, KE $\sim$ 10$^{51}$ erg) is responsible for
the GRB and its afterglows. A broader angle, energetic outflow ($\sim$
1 radian, v/c $\sim$0.1, KE $\sim$10$^{52}$ erg) is responsible for
exploding the star and making the $^{56}$Ni to power the light
curve. A third component of low velocity material exists in the
equatorial plane of those models in which central mass ejection is
blocked by an accretion disk (e.g., collapsars). This component
typically fails to achieve ejection on the first try and falls back to
power a continuing explosion. }
\end{figure}

In the collapsar model \citep{Mac99}, the jet is produced in the near
vicinity of the black hole by neutrinos or MHD processes, while the
large angle ejecta come from the disk wind. In the millisecond
magnetar model, theorists have yet to talk about the two components in
any detail, but one might imagine a large angle component from filling
a cavity with nearly isotropic (thermal?) radiation, while
simultaneously producing a narrower, electromagnetically focused
outflow.  Without better knowledge of the physics of how the
relativistic jet is launched and the efficiency of the disk-wind, it
is not possible to say what fraction of the energy goes into each of
these components, but in all likelihood that fraction varies with the
mass and angular momentum distribution of the presupernova star. It is
also worth keeping in mind that the central engine must not only
maintain a powerful jet for at least the $\sim$10 seconds it takes the
jet to reach the surface of the star, but must hold the direction of
that jet steady to better than 3 degrees \citep{Zha04}. If the jet
wavers by more than that, it becomes contaminated with too many
baryons on the way out to make a GRB.

If the three components can vary independently, and it is hard to see
why they wouldn't, one expects a wide variety of phenomena resulting
from essentially the same central engine. These could include a)
ordinary GRBs [Regions 1 and 2 strongly active]; b) anisotropic
broad-lined supernova without any GRB or bright afterglow
\citep[Region 1 trapped in the star, Region 2 dominant;
e.g., ref.][]{Sod05} c) GRBs with continuing activity after the burst
[Regions 1 and 3 active]; d) XRFs and SN 1998bw [either the cocoon of
an ordinary GRB seen somewhat off axis or an ordinary GRB with higher
baryon loading in the jet]; and others. In this context, SN 1998bw and
SN 2003dh need not be typical of all supernovae associated with GRBs
or XRFs. There would be a continuum of events with the fainter GRBs
and supernovae quite possibly the dominant case in a volume-limited
sample.

\emph{It must be one of the observational goals of the future to gather
a sufficient sample to determine if this is true.} Are GRBs and
core-collapse supernovae a continuum of events with a common central
engine and a smoothly varying parameter - e.g., rotation - underlying
them all. Or are GRBs and supernovae two discrete, different classes
of high energy explosions. Personally, our underlying bias
\citep{Woo05} is that rotation plays little role in most
supernovae. These are the result of (slowly rotating) neutron star
formation and neutrino transport. But ``GRBs'' are a diverse class of
phenomena, overlapping common supernovae in observable
properties. Perhaps rotation powers them all? This is a very old
question, but still a critical one. Just how do massive stars die (and
explode)?

\section{GRBs at High Redshift}

If the properties of GRBs are sensitive to the metallicity as we have
described, then one expects systematic differences in the appearance
of GRBs ``locally'' and at high redshift (where the metallicity is
presumably low). On the average, though perhaps not individually, GRBs
in more metal-deficient regions will come from stars that are more
massive and that have lost less angular momentum. Their disks will
draw from a larger reservoir of matter and, if, as seems reasonable,
the total burst of energy correlates with the total mass accreted, the
GRB will last longer and have more energy. The supernova component may
also be brighter if the disk wind lasts longer and carries more
mass. Bigger helium stars are more tightly bound, however, and harder
to explode \citep{Woo02}, so the supernova could experience
significantly more fall-back. Continuing accretion activity would be
the norm.

It is also possible, in the collapsar model, to have too much angular
momentum \citep{Mac99,Nar01,Lee05}. Angular momentum much in excess of
10$^{17}$ cm$^2$ s$^{-1}$ will lead to a pile up in the disk at such
large radii that neutrino dissipation is negligible. Unable to
dissipate its binding energy, the disk becomes unstable and perhaps
dominated by an outflow, accompanied by very little accretion. Such
behavior has been seen recently in unpublished calculations by Weiqun
Zhang and Andrew MacFadyen. Since the outflow has too little energy to
explode the whole star in one try, a limit cycle may operate (though
this has yet to be followed on the computer). Since angular
momentum increases monotonically outwards in the equator of a GRB
progenitor, material accretes efficiently until a limiting angular
momentum is reached. During that time it maintains a strong jet which
can escape the star and make a GRB. Eventually though, the angular
momentum becomes so large that the disk ceases to be an NDAF
(neutrino-dominated accretion flow) and stagnates. Matter is still
falling in, however, especially from high latitudes. Mixing and shear
will eventually reduce the angular momentum to the point where
accretion can begin again, launch a new jet, and repeat the cycle.
The characteristic time scale would be given by mixing and fall
back. A few hours is reasonable \citep{Mac01}.

This physics is possibly reflected in the observed features of GRB
050904, the most distant GRB discovered \citep{Cum05} and localized
\citep{Kaw05} so far (z = 6.29). This was a long,
multi-peaked, bright burst that lasted well over 205 seconds
\citep{Sak05} and had an equivalent isotropic energy of
0.66 to $3.2 \times 10^{54}$ erg \citep{Tag05,Cus05} and a beaming
corrected energy of 4 to 12 $\times 10^{51}$ erg. Repeated flaring
activity was seen from the burst for 1.5 hours in the rest frame
\citep{Cus05}. While the properties of this burst are not dramatically
different from some others seen closer by, it does lie at an extreme
of energy, duration and variability. It will be very interesting to
see if future bursts at this redshift and higher show these same
characteristics.

\begin{theacknowledgments}
The authors appreciate helpful discussions with and access to the
unpublished calculations of Andrew MacFadyen and Weiqun Zhang.  This
research was supported by NASA (NAG5 12036, MIT 292701, NNG05GG28G,
SWIF03-0047-0037, and NAG5-13700) and the NSF (AST 0206111).  AH is
supported at LANL by DOE contract W-7405-ENG-36 to the Los Alamos
National Laboratory.

\end{theacknowledgments}

\end{document}